\documentclass[aip,rsi,amsmath,amssymb,reprint]{revtex4-1}

\usepackage{graphicx}
\usepackage{amsmath}
\usepackage{amssymb}
\usepackage[titletoc]{appendix}
 
\usepackage{bm}
\usepackage{amsfonts}
\usepackage{epstopdf}
\usepackage{xcolor}
\usepackage{color}
\usepackage{wasysym}
\usepackage{hyperref}
\usepackage{subcaption}

\numberwithin{equation}{section}

\hypersetup{
    colorlinks,    citecolor=blue,    filecolor=blue,    linkcolor=blue,    urlcolor=blue
}

\usepackage{verbatim}

\date{\today}

\begin{document}

\begin{abstract}

We demonstrate and explain the surprising phenomenon of sign reversal in magnetic field amplification by the laser-driven implosion of a structured target. Relativistically intense laser pulses incident on the outer surface of a microtube target consisting of thin opaque shell surrounding a $\mu$m-scale cylindrical void drive an initial ion implosion and later explosion capable of generating and subsequently amplifying strong magnetic fields. While the magnetic field generation is enhanced and spatially smoothed by the application of a kilotesla-level seed field, the sign of the generated field does not always follow the sign of the seed field. One unexpected consequence of the amplification process is a reversal in the sign of the amplified magnetic field when, for example, the target outer cross section is changed from square to circular. Using 2D particle-in-cell simulations, we demonstrate that sign reversal is linked to the stability of the surface magnetic field of opposite sign from the seed which arises at the target inner surface during laser irradiation. The stability of the surface magnetic field and consequently the sign of the final amplified field depends sensitively on the target, laser, and seed magnetic field conditions, which could be leveraged to make laser-driven microtube implosions an attractive platform for the study of magnetic fields in high energy density plasma in regimes where sign reversal either is or is not desired.

\end{abstract}

\title{Sign reversal in magnetic field amplification by relativistic laser-driven microtube implosions}

\author{K. Weichman}
\affiliation{Department of Mechanical and Aerospace Engineering, University of California at San Diego, La Jolla, CA 92093, USA}
\author{M. Murakami}
\affiliation{Institute of Laser Engineering, Osaka University, Suita, Osaka 565-0871, Japan}
\author{A.P.L. Robinson}
\affiliation{Central Laser Facility, STFC Rutherford-Appleton Laboratory, Didcot, OX11 0QX, UK}
\author{A.V. Arefiev}
\affiliation{Department of Mechanical and Aerospace Engineering, University of California at San Diego, La Jolla, CA 92093, USA}

\vskip -2.0cm
\maketitle
\vskip -3.0cm


The emergence of new magnetic field generation techniques~\cite{fujioka2013coil,santos2015coil,gao2016coil,goyon2017coil,ivanov2018zebra}
and structured target fabrication capabilities~\cite{margarone2012nanosphere,cerchez2013grating,jiang2016nanowire,snyder2019structured}
coupled with the continual development of relativistic short pulse lasers~\cite{danson2019lasers}
is rapidly enabling new regimes of magnetized high energy density (HED) physics.
The combination of a strong magnetic field and HED plasma offers the opportunity to observe new magnetization-related phenomena in areas such as laboratory astrophysics~\cite{huntington2015weibel,fiskel2014reconnection,bulanov2015labastro}
and to obtain improvements in applications including inertial fusion energy~\cite{strozzi2012fast_ignition,fujioka2016fast_ignition,sakata2018isochoric}
and ion acceleration~\cite{arefiev2016protons,kuri2018rpa_cp,cheng2019rpa,weichman2020protons}.
In addition, the increasing availability of structured targets has opened up new possibilities for manipulating laser-plasma interaction to achieve desirable goals including enhanced energetic particle production~\cite{klimo2011nanosphere,margarone2012nanosphere,jiang2016nanowire,snyder2019structured}, 
radiation sources~\cite{cerchez2013grating,stark2016gamma,wang2020channel},
and magnetic field generation~\cite{korneev2015snail,korneev2017chiral}.

Concurrently, there is growing interest in scenarios where HED plasma generates or amplifies magnetic fields~\cite{gotchev2009fluxcompression,hohenberger2012flux,shi2020amp,weichman2020surface,murakami2020amp}.
In particular, magnetic field amplification is desirable to extend the experimentally accessible magnetic field beyond what is currently available via vacuum field generation techniques.
One such platform for field amplification is laser-driven implosions.
In the context of 100~$\mu$m-scale implosions driven by ns-duration sub-relativistic-intensity lasers, magnetic fields can be amplified via flux compression~\cite{gotchev2009fluxcompression,hohenberger2012flux}.
However, field amplification is also possible in the implosion of a $\mu$m-scale structured microtube target driven by sub-ps relativistic laser pulses, where the amplification occurs both during and after the ion implosion phase~\cite{murakami2020amp}.

Lasers irradiating a microtube target consisting of a thin opaque shell surrounding a small cylindrical void are capable of driving strong ion acceleration via a two stage process consisting of an initial ion implosion and later explosion~\cite{murakami2018microbubble,murakami2019micro}.
It has recently been demonstrated that this process can also generate strong magnetic fields, which are enhanced by the application of a kilotesla-level seed magneic field~\cite{murakami2020amp}. 
In the proof-of-principle demonstration presented in Ref.~\citenum{murakami2020amp}, the observed strong magnetic field generated within the void had the same direction as the applied seed field.
However, as we will show in this work, a similarly strong magnetic field with peak amplitude in excess of 40 times the seed can also be generated with opposite sign from the seed field.

Magnetic field amplification in microtube targets is a multi-stage process involving field generation by both electron and ion currents. The addition of a seed field also causes the production of a surface magnetic field with opposite sign from the seed at the inner target surface~\cite{weichman2020surface}, which we find under certain conditions can be amplified in lieu of the applied seed. As we will demonstrate in this work, the sign of the strong magnetic field produced by the implosion is influenced by the stability of this surface magnetic field and can be reversed through changes to the target, laser, and seed magnetic field conditions.

\begin{figure*}
    \centering
    \includegraphics[width=0.95\linewidth]{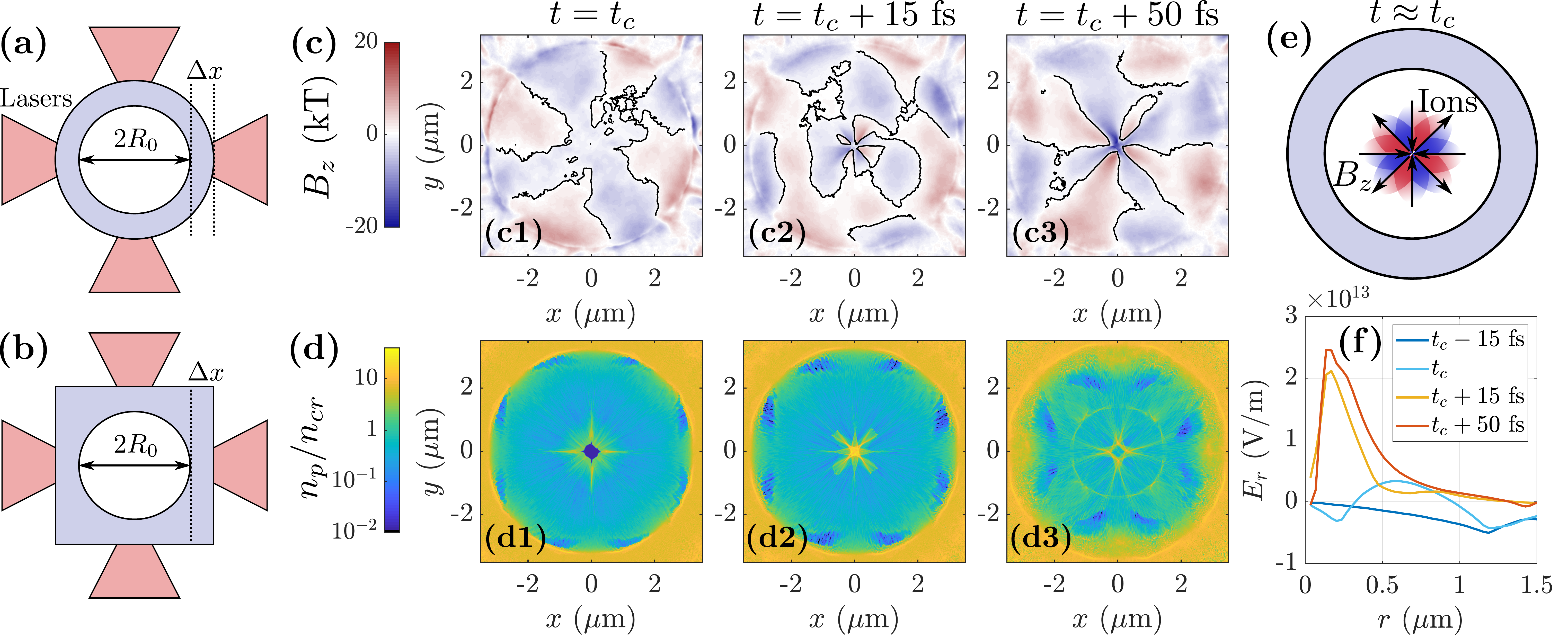}    
    \caption{Magnetic field generation and amplification with $B_{\mathrm{seed}}=0$ in an imploding target. Schematic of target configuration for (a) circular and (b) square outer cross section. (c) Field generation and (d) proton density for circular case with $w_0 = 15$~$\mu$m. The black contours in (c) denote $B_z = 0$.
    (e)~Schematic of $B_z$ generated by imploding ion current. 
    (f)~Azimuthally averaged radial electric field $E_r(r)$ for the circular case with $w_0 = 15$~$\mu$m.
    $t=t_c$ is just before ions reach the center (here, $t_c = 50$~fs). 
    }
    \label{fig:imp_0T}
\end{figure*}

We conduct 2D simulations of an imploding microtube target driven by 4 laser pulses using the open source particle-in-cell code EPOCH~\cite{arber2015epoch}. 
As shown schematically in Fig.~\ref{fig:imp_0T}a,b, 
the target consists of a thin fully ionized plastic (CH) shell which is either completely cylindrical or has a square outer cross section with the same central cylindrical hole. The minimum thickness of this shell is nominally 3~$\mu$m with a 3~$\mu$m radius hole. We nominally apply a seed magnetic field in the direction out of the simulation plane ($z$) of $B_{\mathrm{seed}}=3$~kT. The 4 laser pulses are spatially and temporally Gaussian with a peak intensity of $10^{21}$~W/cm$^2$, a 25~fs FWHM pulse duration, and a varying spot size $w_0$.
Additional details of the simulation setup are given in Table~\ref{table:2DPIC_implosion}.

\begin{table}
\centering
\begin{tabular}{ |l|l| }
  \hline
  \multicolumn{2}{|l|}{\textbf{Laser parameters} }\\
  \hline
  Wavelength & $\lambda_0=0.8$ $\mu$m \\
  Peak intensity & $1 \times 10^{21}$ W/cm$^2$ \\
  Duration (Gaussian, electric field FWHM) & $25$ fs\\
  Spot size (Gaussian, electric field FWHM) & $w_0 = 15$ $\mu$m\\
  Laser polarization & $x$ or $y$ \\
  \hline \hline
  \multicolumn{2}{|l|}{\textbf{Other parameters} }\\
  \hline
  Seed magnetic field ($\mathbf{B}=B_{\mathrm{seed}} \mathbf{\hat{z}}$) & $B_{\mathrm{seed}}=3$ kT \\  
  Target inner radius & $R_0 = 3$ $\mu$m\\
  Minimum target thickness & $\Delta x = 3$ $\mu$m\\
  Peak electron density & $n_e = 50\;n_{cr}$ \\  
  Spatial resolution & 100 cells/$\lambda_0$ \\
  Macroparticles per cell, electron &  200 \\
  Macroparticles per cell, ion &  100 \\  
  Size of simulation box ($x \times y$, $\mu$m) & $48 \times 48$ \\
  Time interval for averaging $B_z$ in figures & 5 fs \\
  \hline \hline
  \multicolumn{2}{|l|}{\textbf{Time reference}} \\
  \hline
  Time when peak of laser would reach void & $t = 0$ \\
  Time just before ions reach center (varies) & $t=t_c$ \\
  \hline
   \end{tabular}
  \caption{
  Implosion simulation parameters. 
  The inner target cross section is circular with either a circular or square outer cross section. The initial plasma temperature is set as zero and the target surface is sharp (no preplasma). $t_c$ is measured to within 5~fs.}
  \label{table:2DPIC_implosion}
\end{table}

The essential dynamics of the implosion are as follows~\cite{murakami2018microbubble} and are qualitatively unchanged by the addition of the seed magnetic field.
The laser pulse interacting with the outer target surface generates hot electrons which stream into the inner target void and drive a strong ion implosion towards the target center (Fig.~\ref{fig:imp_0T}d).
This implosion is driven by the radially inward electric field associated with the net excess of electron charge within the void ($t \lesssim t_c$ in Fig.~\ref{fig:imp_0T}f, where $t_c$ is the time ions first reach the target center). 
After ions arrive at the target center, the radial electric field is reversed and acts to re-accelerate ions outward during a subsequent explosion phase ($t \gtrsim t_c$ in Fig.~\ref{fig:imp_0T}f).
This process can generate high plasma density at the target center and produce high ion energy from the explosion~\cite{murakami2018microbubble,murakami2019micro}.

Even with no initial seed magnetic field ($B_\mathrm{seed}=0$), the departure from cylindrical symmetry introduced by the 4 laser spots produces a strong $\pm z$-directed magnetic field. 
This magnetic field initially has two components, a hot-electron-generated component near the edge of the target void which is visible early in time (near $r= 3\;\mu$m in Fig.~\ref{fig:imp_0T}c1), and a later ion-generated component near the target center ($r\lesssim 1\;\mu$m in Fig.~\ref{fig:imp_0T}c2). The ion-generated magnetic field is driven by the strong spatial non-uniformity in the imploding ion density (Fig.~\ref{fig:imp_0T}d), which produces the 8-lobed magnetic field profile shown schematically in Fig.~\ref{fig:imp_0T}e.
During the subsequent explosion ($E_r>0$) phase, the net flow of electrons towards the target center produces a net $-E_r \times B_z$-directed current, and amplifies the ion-generated magnetic field.
In the case of $B_{\mathrm{seed}}=0$, this process results in an amplified magnetic field profile with sub-$\mu$m-scale structure including both positive and negative lobes (Fig.~\ref{fig:imp_0T}c3).

\begin{figure*}
    \centering
    \includegraphics[width=0.95\linewidth]{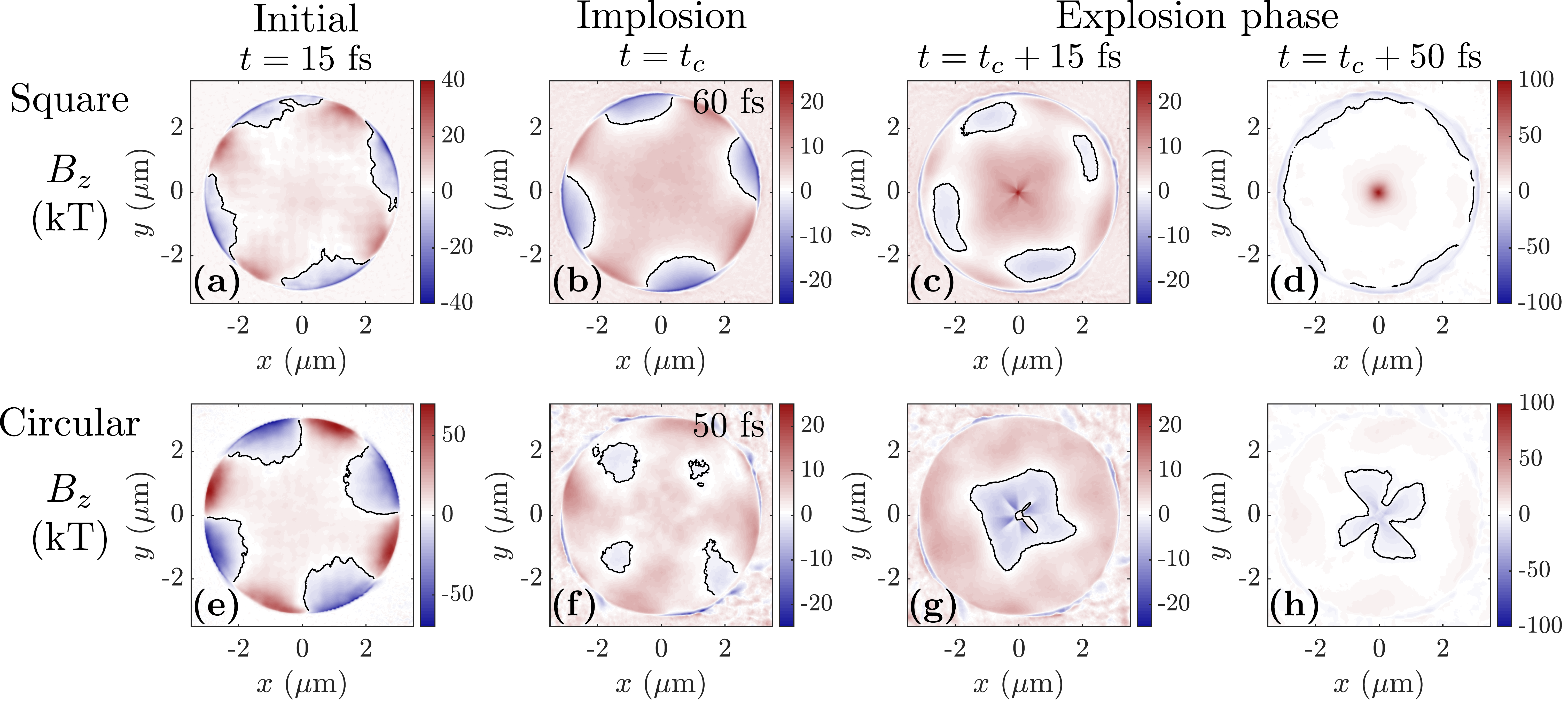}
    \caption{Time history of magnetic field amplification with $w_0=15$~$\mu$m and $B_{\mathrm{seed}} = 3$~kT. 
    (a)-(d)~Square outer cross section target. (e)-(h)~Circular outer cross section target.
    $t_c$ is just before ions cross through the center of the target.
    }
    \label{fig:imp_shape}
\end{figure*}

The addition of $B_\mathrm{seed}= 3$~kT leads to the generation of a field which is stronger and more spatially uniform than the field generated with $B_\mathrm{seed}=0$ (e.g. Fig.~\ref{fig:imp_shape}d vs Fig.~\ref{fig:imp_0T}c), but otherwise has little effect on the implosion process. 
With the seed field, the magnetic field can be amplified by the imploding ions (within 10-30 fs of $t_c$) in addition to the later amplification by $E \times B$ electron current~\cite{murakami2020amp}. 
Importantly, the magnetic field which is amplified by this process is the \textit{locally present} magnetic field.
The locally present field is the seed field in part of the parameter space, 
for example under the conditions shown in Fig.~\ref{fig:imp_shape}a-d, and in Ref.~\cite{murakami2020amp}.
However, the application of the the seed magnetic field also triggers the production of a strong magnetic field with opposite sign at the inner target surface~\cite{weichman2020surface}. 
This surface-generated magnetic field can under other conditions be amplified in lieu of the seed, for example in Fig.~\ref{fig:imp_shape}e-h, reversing the sign of the amplified field.

\begin{figure}
    \centering
    \includegraphics[width=0.8\linewidth]{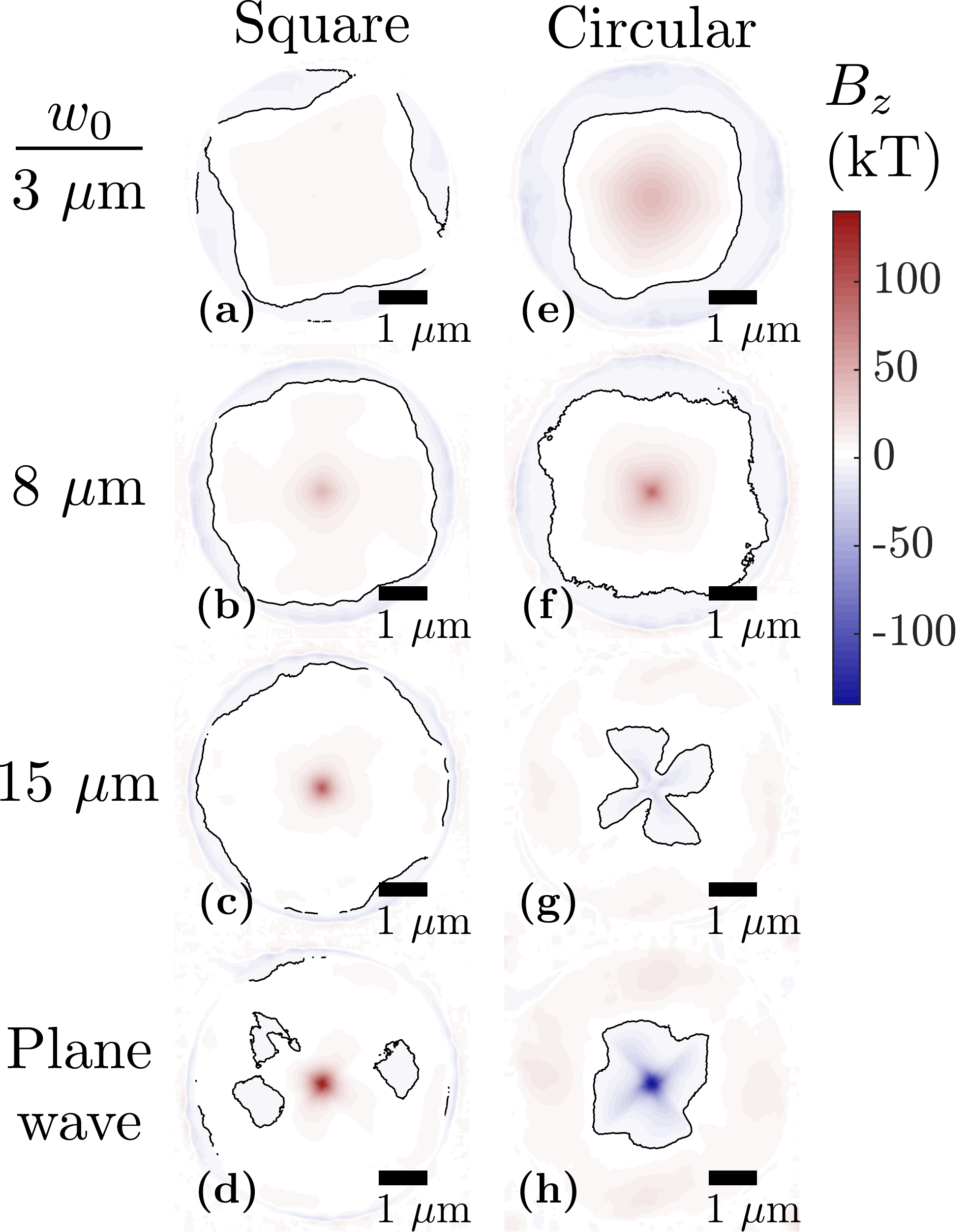}
    \caption{Comparison of magnetic field produced by imploding target with different target outer shapes (see Figs.~\ref{fig:imp_0T}a,b) and laser spot size with $B_{\mathrm{seed}} = 3$~kT. The magnetic field is shown well after the implosion phase ($t=t_c+50$~fs). The black contours denote $B_z = 0$.
    }
    \label{fig:imp_cases}
\end{figure}

Surface magnetic field generation and the processes leading to sign reversal lead to a significant dependence of the amplified magnetic field on the target, laser, and seed magnetic field conditions.
A surprising consequence of this parameter dependence is shown in Fig.~\ref{fig:imp_cases}, where we show, for example, that the sign of the amplified magnetic field can in some cases be reversed by changing the outer cross section of the target from square to circular.
The amplified magnetic field can even obtain the same magnitude in spite of the sign reversal (for example, the greater than 40-fold amplification relative to the seed shown in Cases d and h in Fig.~\ref{fig:imp_cases}). 
This is distinctly different than what would be expected from the geometric flux compression observed in 100~$\mu$m-scale implosions driven by sub-relativistic laser pulses~\cite{gotchev2009fluxcompression,hohenberger2012flux}, and may be a unique feature of magnetic field amplification in $\mu$m-scale relativistic laser-driven implosions. 

\begin{figure}
    \centering
    \includegraphics[width=0.85\linewidth]{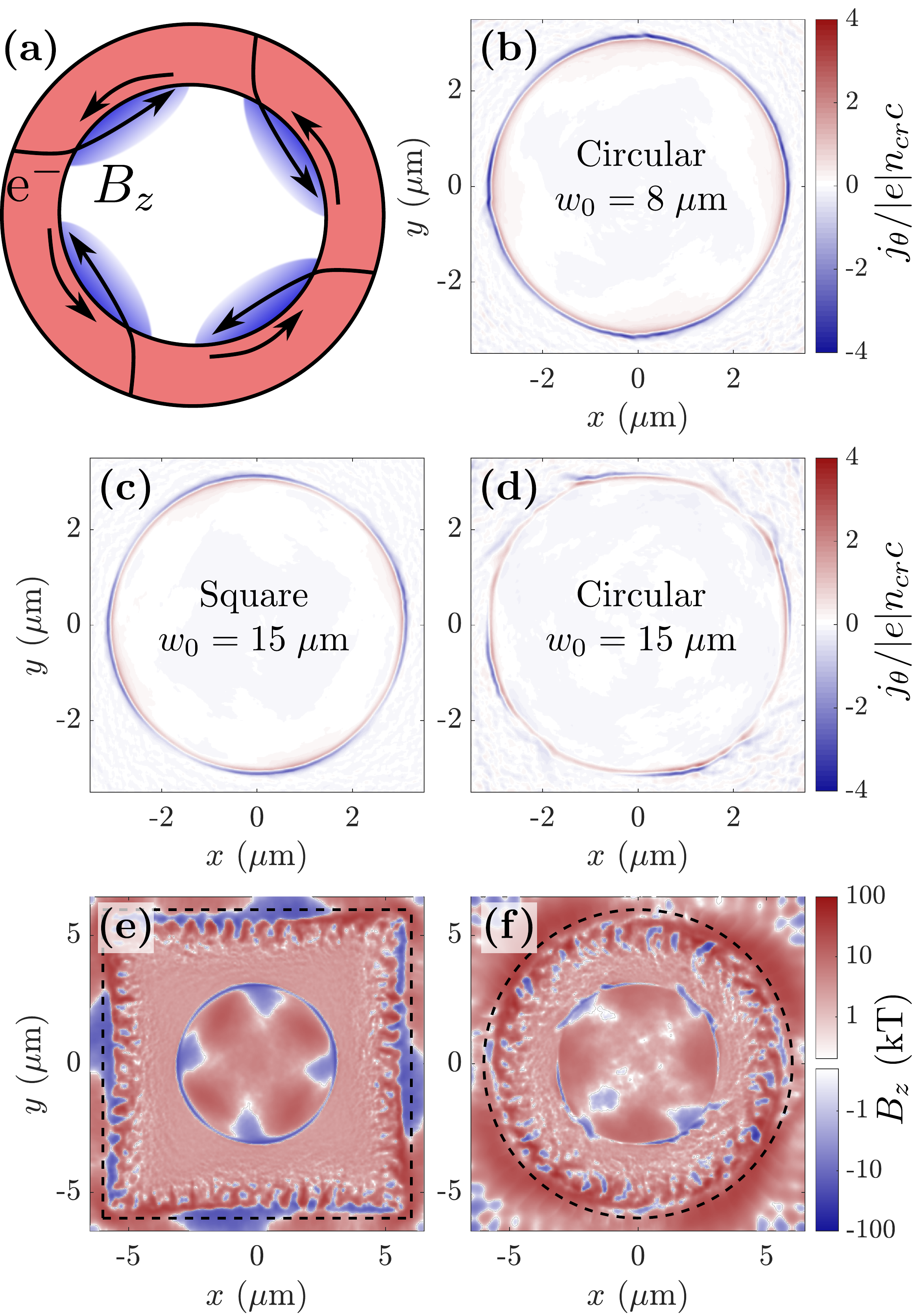}
    \caption{Formation and disruption of surface magnetic field. (a) Conceptual diagram of surface magnetic field generation with opposite sign from the applied seed. 
    (b)-(d) Azimuthal current density $j_\theta$ at $t=50$~fs for (b) circular target with 8~$\mu$m laser spot, (d) square target with 15~$\mu$m spot, (d) circular target with 15~$\mu$m spot. (e)-(f) Magnetic field in the 15~$\mu$m spot cases at $t=45$~fs with (e) square and (f) circular outer cross section. Dashed lines indicate the initial target outer surface.
    }
    \label{fig:imp_disrupt}
\end{figure}

Whether the seed or the surface-generated magnetic field is amplified depends on the stability of the magnetic field at the target surface around the time ions first pass through the target center.
The surface magnetic field arises due to the cyclotron rotation of laser-heated electrons transiting radially through the target~\cite{weichman2020surface} (shown conceptually in Fig.~\ref{fig:imp_disrupt}a). The current associated with the cyclotron rotation of these hot electrons and the compensating return current in the target create a double current layer near the surface (e.g. Fig.~\ref{fig:imp_disrupt}b), which produces the surface magnetic field.
The surface-generated field competes with and can also suppress the electron-associated field~\cite{weichman2020surface}, and is visible in the dominance of the $-z$-directed field close to (within $\sim 0.1$~$\mu$m of) the target inner surface in Figs.~\ref{fig:imp_shape}a,e.

The surface current generation process is disrupted if the magnetic field within the  target bulk changes substantially during the implosion. The four laser pulses driving the implosion seed a periodic structure in the plasma density and magnetic field at the outer target surface. These initial perturbations combined with the ongoing streaming of hot electrons through the relatively cold target produce a growing filamentary magnetic field structure within the target itself (Figs.~\ref{fig:imp_disrupt}e,f). 
The penetration of these filaments deep into the target disrupts the surface current and the surface magnetic field (for example, Figs.~\ref{fig:imp_disrupt}d,f). Whether the filaments are able to grow through the full target thickness and disrupt the surface magnetic field in time to affect the magnetic field amplification depends on the target, laser, and seed magnetic field parameters. The higher the population of hot electrons and the weaker the applied magnetic field, the faster the filaments grow. 

In the context of the scan over the laser spot size and outer target shape shown in Fig.~\ref{fig:imp_cases}, we observe a larger population of recirculating hot electrons in the cases where the magnetic field within the target is disrupted before the ions reach the target center (Cases g-h) than in the other cases. More hot electrons are initially produced in these cases due to the large laser spot and the departure of the laser from normal incidence provided by the circular outer cross section (which allows for more efficient electron production from a sharp interface~\cite{gibbon1992absorption}). 

The impact of this difference in the recirculating hot electron population on the magnetic field amplification process can be seen for example in the comparison of the circular and square outer cross section cases with $w_0 = 15\;\mu$m shown in Figs.~\ref{fig:imp_shape} and~\ref{fig:imp_disrupt}. 
In the square case, the magnetic field within the target bulk remains mostly unperturbed (Fig.~\ref{fig:imp_disrupt}e) and the surface current is stable (Fig.~\ref{fig:imp_disrupt}c). The magnetic field near the target center maintains the same sign as the applied seed throughout the implosion and the final amplified magnetic field is positive (Fig.~\ref{fig:imp_shape}a-d).
In contrast, in the circular case, the magnetic field filaments penetrate through the full target thickness by around $t=45$~fs (Fig.~\ref{fig:imp_disrupt}f), disrupting the surface current (Fig.~\ref{fig:imp_disrupt}d). Before this time, the magnetic field in the void is qualitatively similar between the circular and square cases, albeit with somewhat different magnitude (Figs.~\ref{fig:imp_shape}a,e). After this time, however, the surface magnetic field in the circular case is broken up and regions of negative magnetic field are pushed into the target center (Fig.~\ref{fig:imp_shape}f). The timing of the disruption of the surface magnetic field is such that this negative, originally surface-generated field is present in the center during the explosion phase and is amplified in lieu of the seed field (Fig.~\ref{fig:imp_shape}g-h).

\begin{figure}
    \centering
    \includegraphics[width=0.75\linewidth]{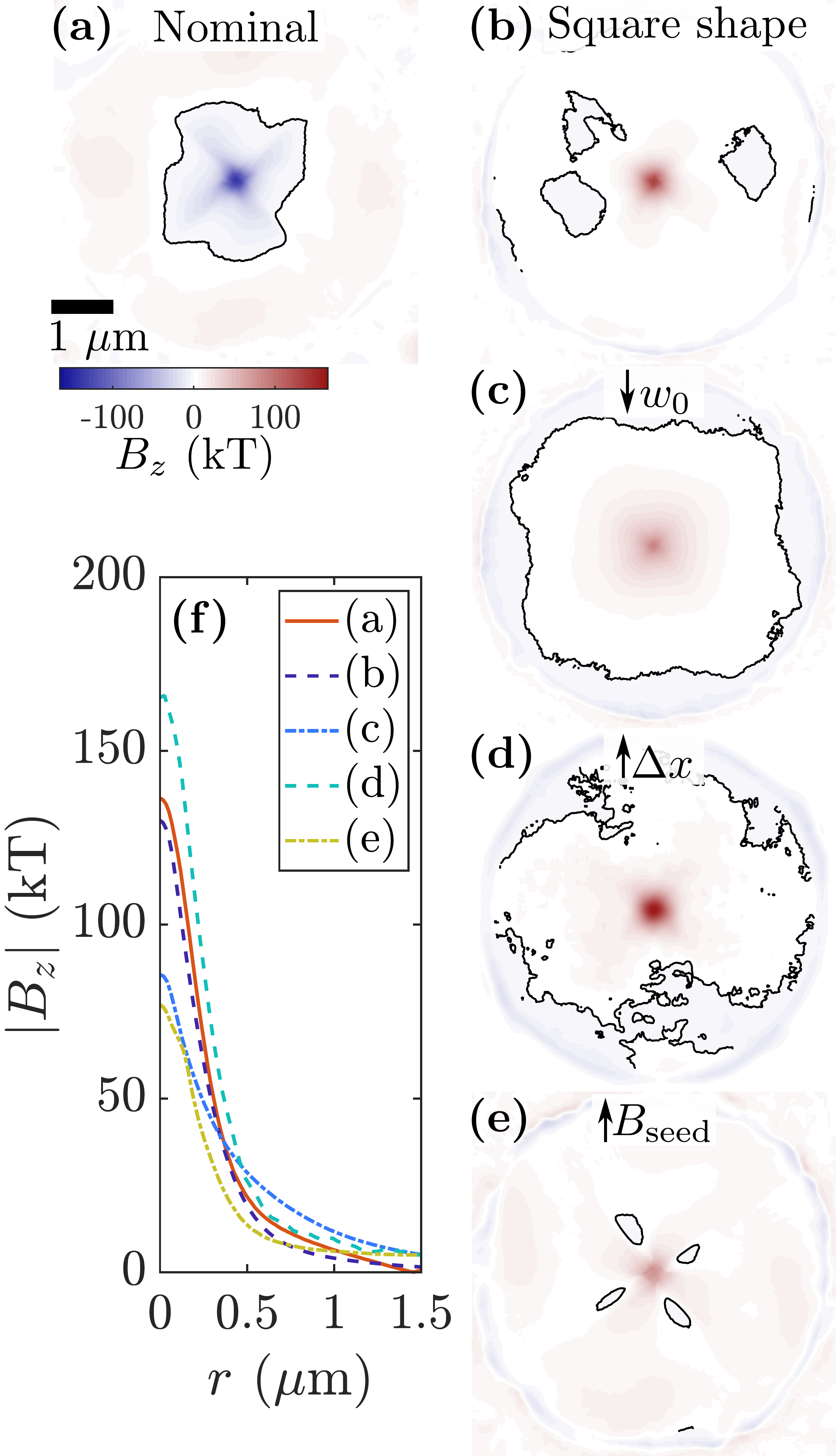}
    \caption{
    Dependence of magnetic field sign reversal on key parameters.
    (a) Circular plane wave case, with other parameters as given in Table~\ref{table:2DPIC_implosion}. (b)-(e) denote changes relative to the case given in (a), with (b) square outer cross section, (c) $8\;\mu$m laser spot size, (d) $6\;\mu$m thick target, and (e) $B_\mathrm{seed}=6$~kT.
    (f)~Azimuthally averaged magnetic field $|B_z(r)|$ corresponding to (a)-(e).
    The time shown is $t=t_c+50$~fs. Black contours in (a)-(e) denote $B_z=0$. 
    }
    \label{fig:imp_summary}
\end{figure}

In addition to the target outer shape and laser spot size, the sign reversal of the amplified magnetic field relative to the seed can depend on a number of other target, laser, and seed magnetic field parameters.
In principle, any parameter which affects the timing of the implosion or the growth of the magnetic field filaments could affect the sign of the final generated field.
To further illustrate this sensitivity, we have additionally scanned over the target thickness and seed magnetic field strength.
Figure~\ref{fig:imp_summary} summarizes how these parameters, as well as the target shape and laser spot size, affect the sign reversal of the amplified magnetic field.

First, we consider the effect of the target thickness. We have conducted additional simulations with a spatially plane wave laser pulse with the same FWHM pulse duration as given in Table~\ref{table:2DPIC_implosion} and a $\sin^2$ temporal shape in $|E|$ (to reduce the computational cost), scanning over the minimum target thickness $\Delta x$. For this parameter scan, we specifically consider the effect of $\Delta x$ on the sign reversal in a target with outer circular cross section. The change of the temporal shape of the laser pulse from Gaussian to $\sin^2$ has negligible effect on the magnetic field observed in the target and the final magnetic field profile in the $\Delta x=3\;\mu$m case is nearly identical to the result given in Fig.~\ref{fig:imp_cases}h.

We find that decreasing the target thickness relative to the nominal case (e.g. decreasing $\Delta x$ from 3~$\mu$m to 1~$\mu$m) has no effect on the sign of the magnetic field. If, however, the target thickness is increased (e.g. $\Delta x=6\;\mu$m), the magnetic filaments are unable to penetrate through the full target thickness before the implosion, the surface magnetic field is stable, and the final generated field has the same sign as the applied seed (Fig.~\ref{fig:imp_summary}d).

Second, we consider the effect of the seed magnetic field strength. We return to the conditions given in Table~\ref{table:2DPIC_implosion} (temporally Gaussian laser pulse with $\Delta x =3\;\mu$m).
We observe that increasing $B_\mathrm{seed}$ from 3~kT (the nominal case) to 6~kT changes the sign of the final generated field (Fig.~\ref{fig:imp_summary}e). In the 6~kT case, the increased $B_\mathrm{seed}$ prevents the magnetic field filaments from penetrating deep into the target and allows the surface-generated magnetic field to be stable.

This dependence of the sign of the amplified magnetic field on $B_\mathrm{seed}$ was previously seen in Ref.~\citenum{murakami2020amp}, in simulations where the laser-plasma interaction was replaced by an initial distribution of hot electrons. 
The observed sign reversal of the magnetic field in the hot electron case may also be attributable to the destabilization of the surface-generated magnetic field. However, the inclusion of hot electrons throughout the target instead of their generation at the surface, as well as the lack of laser-imposed perturbations likely changes the dynamics of the surface field breakup.

In summary, we have demonstrated that magnetic field amplification in imploding microtube targets can produce magnetic fields with amplitude in excess of 40~times an applied seed field with a polarity that either matches the seed or is opposite to it. 
The potential to generate a strong magnetic field with either sign is related to the ability of the microtube target to amplify the locally present magnetic field in the void and the production of a strong magnetic field at the inner target surface with opposite sign to the seed. Whether the applied seed field or the surface field is amplified depends on the stability of the surface current, which is determined by the growth of laser-seeded magnetic filaments within the target.
Consequently, we find that the sign of the generated magnetic field can be reversed by changes to the target, laser, and seed magnetic field configuration.
This ability to reverse the magnetic field with only small changes to the experimental configuration could make this system an interesting platform for future study of the generation and effects of strong magnetic fields in high energy density plasma.

This research was supported by the DOE Office of Science under Grant No. DE-SC0018312.
Particle-in-cell simulations were performed using EPOCH~\cite{arber2015epoch}, developed under UK EPSRC Grant Nos. EP/G054940, EP/G055165, and EP/G056803.
This work used HPC resources of the Texas Advanced Computing Center (TACC) at the University of Texas at Austin and the Extreme Science and Engineering Discovery Environment (XSEDE)~\cite{towns2014xsede}, which is supported by National Science Foundation grant number ACI-1548562.
Data collaboration was supported by the SeedMe2 project~\cite{chourasia2017seedme} (http://dibbs.seedme.org).

The data that support the findings of this study are available from the corresponding author upon reasonable request.

%

\end{document}